# Redundancy of Stereoscopic Images: Experimental Evaluation

***L. P. Yaroslavsky*[1], *J. Campos*[2], *M. Espínola*[2], *I. Ideses*[1]**

[1] Dept. Interdisciplinary Studies. Faculty of Engineering. Tel Aviv University. Tel Aviv, Ramat Aviv 69978, Israel
[2] Dept. Física, Universidad Autónoma de Barcelona, 08193 Bellaterra, Spain
[1]**e-mail**: yaro@eng.tau.ac.il, [2]Juan.Campos@uab.es

**Abstract:** With the recent advancement in visualization devices over the last years, we are seeing a growing market for stereoscopic content. In order to convey 3D content by means of stereoscopic displays, one needs to transmit and display at least 2 points of view of the video content. This has profound implications on the resources required to transmit the content, as well as demands on the complexity of the visualization system. It is known that stereoscopic images are redundant, which may prove useful for compression and may have positive effect on the construction of the visualization device. In this paper we describe an experimental evaluation of data redundancy in color stereoscopic images. In the experiments with computer generated and real life and test stereo images, several observers visually tested the stereopsis threshold and accuracy of parallax measuring in anaglyphs and stereograms as functions of the blur degree of one of two stereo images and color saturation threshold in one of two stereo images for which full color 3D perception with no visible color degradations is maintained. The experiments support a theoretical estimate that one has to add, to data required to reproduce one of two stereoscopic images, only several percents of that amount of data in order to achieve stereoscopic perception.

## References and links

## 1. Introduction

As digital imaging techniques gain popularity and acceptance, rendering the race for increased spatial resolution marginal, the field of 3D imaging and visualization remains one of the last uncharted frontiers. Interest in 3D visualization and 3D content has been constantly growing as imaging devices were developed, even though a killer application for 3D visualization is yet to be developed. It is obvious, however, that future imaging devices will not be confined to 2D, but rather will capture 3D images to be displayed either in 2D or 3D at will.

There are many different ways to display 3D images [1,2]. Most methods for 3D display are based on stereopsis which is one of the most important visual mechanisms of 3D vision ([6]). In stereoscopic displays, two images of the 3D scene obtained from two viewing positions are projected separately on left and right eyes of the viewer thus enabling 3D perception through the stereoscopic vision mechanism.

The most straightforward way to display stereoscopic images is using stereoscopes. Stereoscopic and, in particular, autostereoscopic displays are considered the high-end solution for 3D visualization [2]. These devices exhibit excellent stereo perception in color without the need for viewing aids. Overall cost and viewing area limitations, however, inhibit the market share of such devices, rendering their market acceptance marginal.

A simple and inexpensive alternative are anaglyphs. Anaglyphs are artificial color images in which one color component represents one of two stereoscopic images and one or two other components represent the second stereoscopic image. For viewing anaglyphs, special red-blue or red-green glasses are needed. Anaglyph based 3D projection is considered the lowest cost stereoscopic projection technique. This method can be implemented on standard monitors without any modification.

An important issue in the development of 3D visualization and visual communication methods is efficient coding of the data required for generation and display of stereoscopic images. It is well known that stereoscopic images are very redundant in their information content, however, it is much less known how substantial this redundancy is and how it can be used for efficient coding of stereoscopic images for storage and transmission. In this paper we address this problem from a visual communication perspective and experimentally investigate the tolerance of 3D stereoscopic vision to loss of resolution and color saturation of one of two stereoscopic images when regular high quality computer color displays are used to display images.

Three sets of experiments are reported in the paper: (i) evaluation of the threshold of stereopsis to loss of resolution in one of two stereoscopic images due to low pass filtering, (ii) evaluation of the accuracy of visual measuring of parallax as function of the degree of image blur and (iii) tolerance of 3D color visual perception to reduced color saturation of one of two stereo images. For image blur, two methods were used: image blurring by means of an approximated ideal low pass filter and image blurring by means of low pass filters used in wavelet image decomposition. The latter was motivated by the fact that modern image compression techniques are based on wavelet transform domain coding.

The paper is organized as follows. In Sect. 2 we discuss theoretical considerations that provide an estimate of the order of magnitude of the redundancy of stereoscopic images. In Sect. 3 we describe the experimental methodology and setup. In Sect. 4, experimental results obtained with image blurring are given. In Sect. 5, experimental results on color saturation threshold are provided. In the Conclusion, we summarize and discuss obtained results.

## 2. Redundancy of stereoscopic images: qualitative evaluation

One can make an estimation of the redundancy of stereoscopic images using the following rationale.

From the informational point of view, two images of the same scene that form a stereo pair are equivalent to one of the images and a depth map of the scene. Indeed, from two images of the stereo pair, one can build the depth map, and, vice-versa, one can build a stereo pair from one image and the depth map. Therefore, the increase of the signal volume the second image of the stereo pair adds to that of one image is equal to the signal volume that corresponds to the depth map.

The number of depth gradations resolved by vision is of the same order of magnitude as the number of resolved image gray levels. Therefore, signal volume increment due to the depth map is basically determined by the number of depth map independent samples.

Every sample of the depth map can be found by localizing corresponding fragments of two images of the stereo pair and measuring parallax between them. All technical devices that measure depth from stereo work in this way, and it is only natural to assume the same mechanism for stereoscopic vision. The number of independent measurements of the depth map is obviously the ratio of the image area to the minimal area of the fragments of one image that can be reliably localized on another image. It is also obvious, that it is, generally, not possible to reliably localize one pixel of one image on another image. For reliable localization, image fragments should contain several pixels. Therefore, the number of independent samples of the depth map will be, correspondingly, several times lower than the number of image pixels, and the increment of the signal volume that corresponds to the depth map will be several times lower than the signal volume of one image. For instance, if the reliable size of the localized fragment is 2x2 pixels, it will be four times lower, for 3x3 fragments, it will be 9 times lower, and so on. Practical experience tells that, for reliable localization of fragments of one image on another image, the fragment size should usually exceed the area of 8x8 to 10x10 pixels.

Therefore one can hypothesize that the signal volume increment associated with the depth map may amount to percents or even fractions of a percent of the signal volume of one image and that this limitation of the resolution of depth maps acquired from stereo images is true for any device for extraction of depth information from stereo images, including human stereoscopic vision. This rationale motivates experimental verification of this hypothesis. In what follows, we present experimental data in its support.

## 3. Description of experiment methodology and general experimental setup.

The experiments were aimed at evaluation, from a visual communication perspective, of 3D color vision tolerance to blur and to loss of color saturation of one of two stereoscopic images. They were conducted with artificial computer generated stereo images as well as with real life images. As the goal of the experiments was an evaluation of the vision tolerance to image distortions when regular high quality computer monitors were used for image display with no special adjustment for viewing conditions, standard (default) monitor settings of the monitor primaries and of the monitor calibrated tone reproduction curves were used. In the experiments, images were displayed on regular computer LCD and CRT high quality monitors for visual observation using either a stereoscope or anaglyphs. Specifically, LCD monitor ACER AL-2021 with resolution 1600x1200 pixels and CRT monitor MAG 786PF with resolution 1024x768 pixels were used. Observations were carried out in darkened rooms with viewing distance 40-60 cm, for CRT monitors, the white color temperature was set to 9300º Kelvin.

As test images, the following images were used:

- Grayscale random dots images (see an example in Fig. 1, a)
- Grayscale random patches images (Fig. 1, b)
- Grayscale texture images (Fig. 1, c, d )
- Color random patches images (Fig. 1, e, f)
- Real-life stereoscopic images (see an example in Fig. 1 g), h), right and left, respectively, for crossed eyes viewing)

Computer generated test images were of spatial resolution of 512x512 pixels , real-life stereoscopic images were of the comparable size.

For each of the test images except the real-life stereo ones, its artificial stereo pair was artificially synthesized using, as a depth map, impulses of square or round shapes of different dimensions (Fig. 2, a,b). Discrete sinc interpolation was used for image resampling needed for generating images with artificially introduced parallax ([5)].Fig. 3 shows examples of stereo images generated in this way.

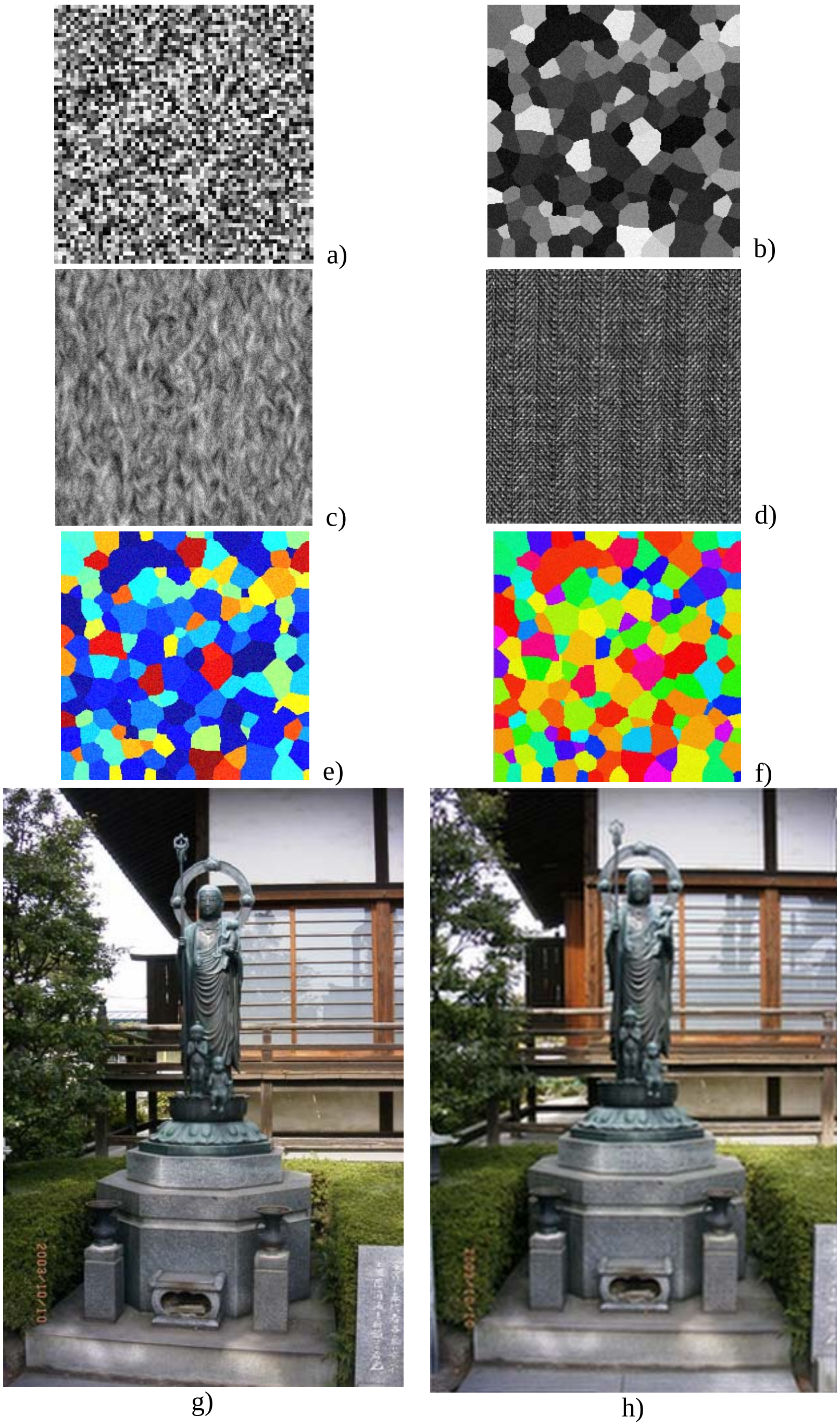

Fig. 1. Test images used in the experiments. Image h) exemplifies low pass filtering to 1/4 of the image base band. Note that when images g) and h) are fused using stereoscope or by crossed eyes, a 3-D sharp image can be seen

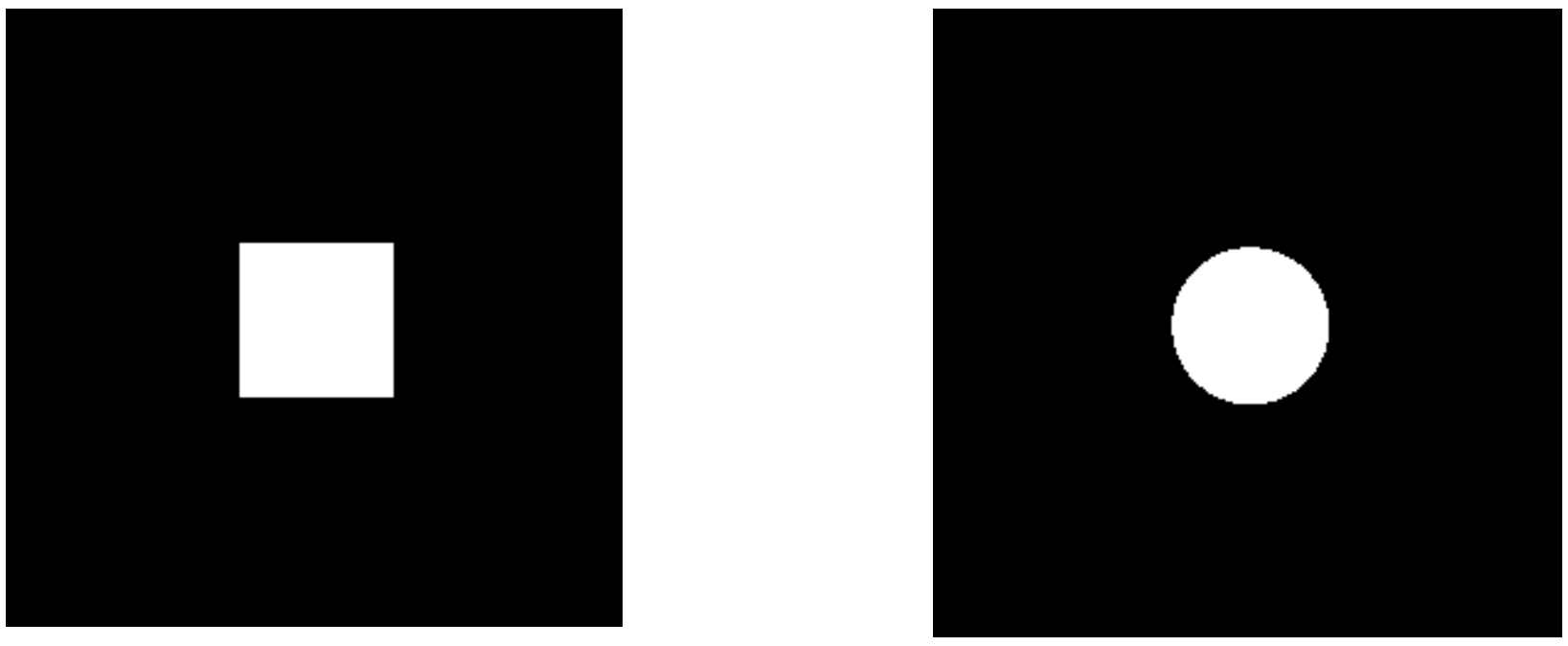

Fig. 2. Depth maps used in the experiments

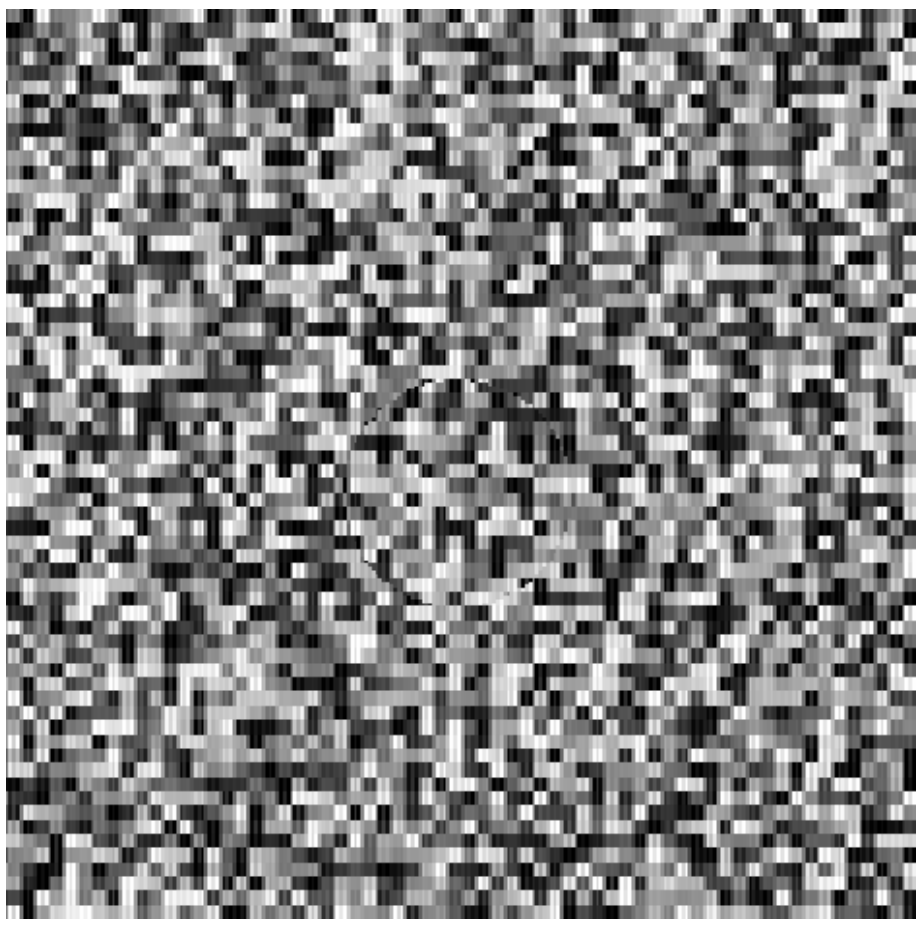

a) Right eye

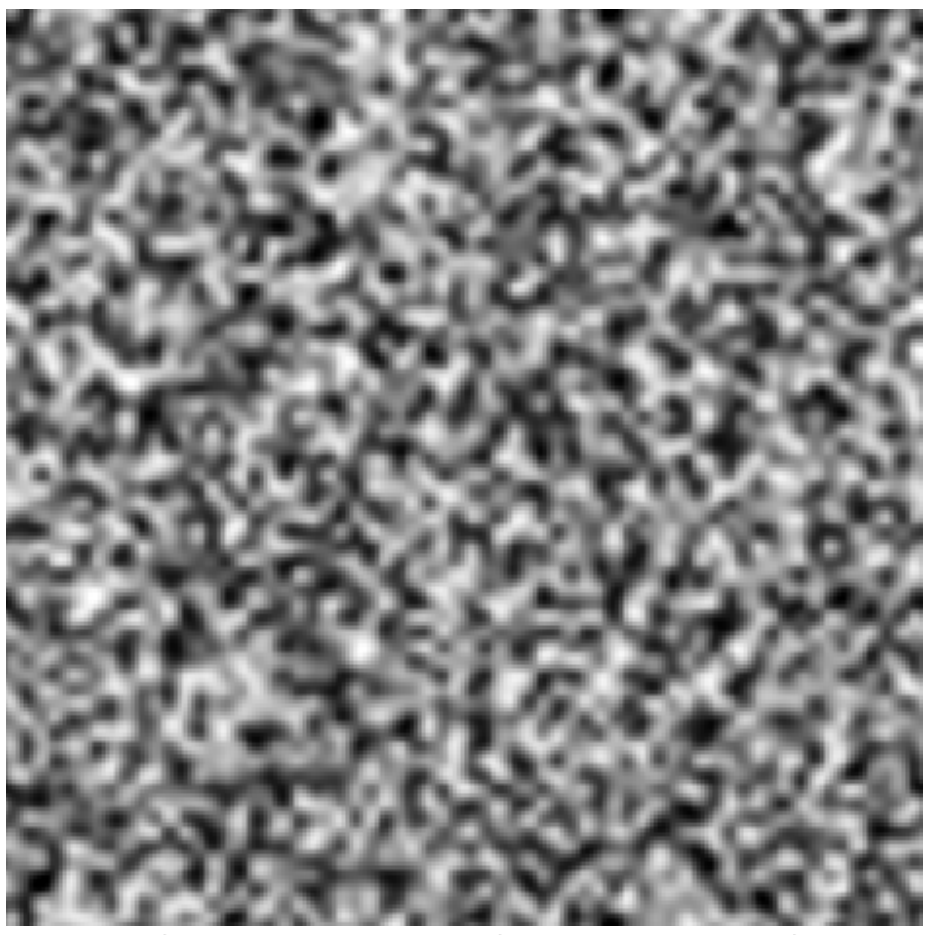

b) Left eye, low pass filtered to ¼ of the base band

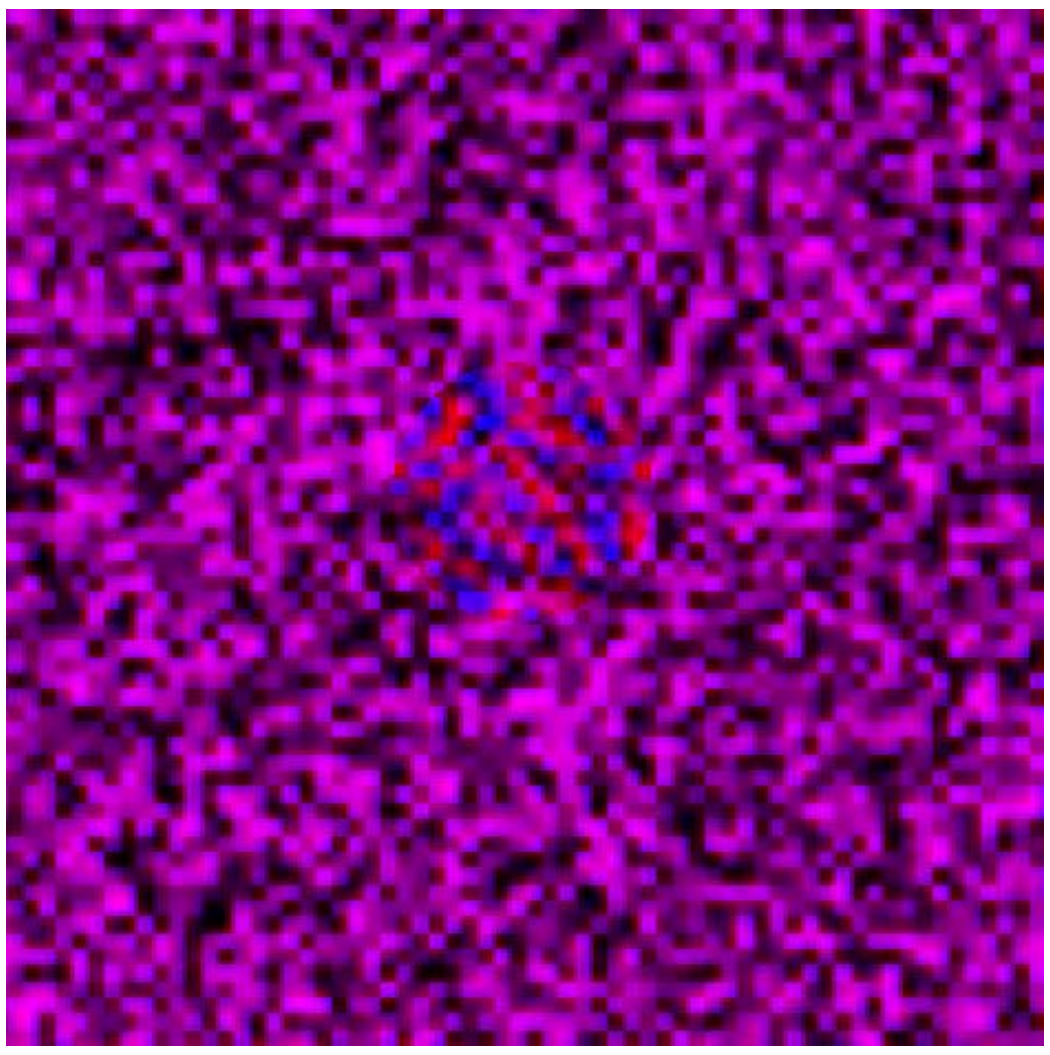

c)

Fig. 3. Examples of computer generated random dot patterns used in the experiments: a), b) – right and left images for stereoscope or crossed eyes view, c) –anaglyph (left eye–red, right eye- blue). Right images exemplify low pass filtering to ¼ of the image baseband. Note that when images are fused using stereoscope or crossed eyes and, correspondingly, color red-blue glasses, a test circle 3-D target can be seen without visible losses of the image resolution

For each particular experiment, the center of the impulse was randomly placed within the image area so as to exclude the viewer's adaptation. In the experiments on the stereopsis threshold, the tested values of the depth map scale parameter that determines image parallax were randomly reshuffled for the same purpose. In addition, before displaying each next image, the display was kept blank for several seconds to secure the absence of a cross-talk between individual experiments.

All experiments were carried out with several viewers. For each viewer and for each of the measured parameters, statistics were accumulated over several tens of realizations of random depth map impulse positions. The obtained data were statistically averaged using arithmetic mean and median averaging and standard deviation of the measured data was determined. In addition, time delay between the prompt to viewer to act and the viewer's response was measured in all experiments in order to obtain supplementary data to the basic ones.

## 4. Tolerance of 3D visual perception to image blur

Image blur in both stereoscopic images is frequently regarded as a depth cue (see, for instance, [4]). In this paper, we consider influence on stereopsis of blur introduced to one of two stereoscopic images. Perhaps one of the first observations of the phenomenon of the tolerance of stereoscopic vision to blur of one of the stereo images was made by Bela Julesz [5], who however did not mention any quantitative measures of this phenomenon. The first experiments reported in the literature of decimation and subsequent interpolation of one of two stereoscopic images were described in [6]). Results obtained there are summarized in Fig. 4 that shows the root mean squared error of parallax measurements on a training stereo air photograph analyzed by a professional human operator using a stereocomparator when one of the images was decimated with different decimation factors and then bilinearly interpolated back to the initial size. The data plotted in Fig. 4 were obtained by averaging of the parallax measuring errors over 31 randomly selected image fragments. Decimation and interpolation were carried out in a computer and the operator was working with computer-generated images of a good photographic quality. Decimation factor zero corresponds to data obtained for initial photos (not computer printouts).

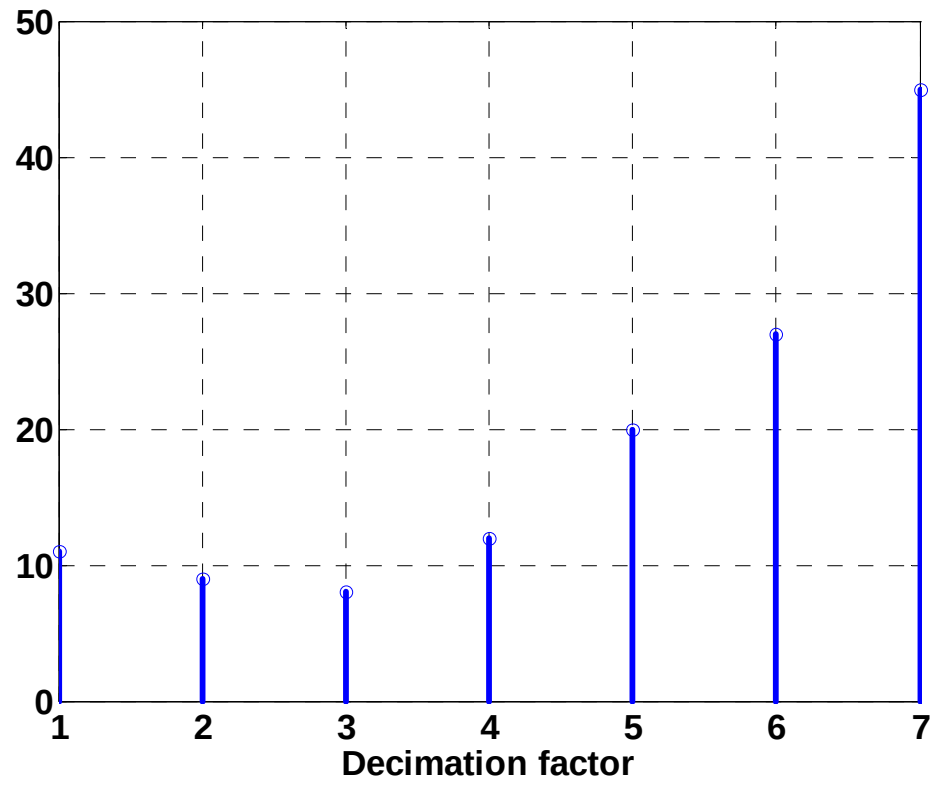

Fig. 4 Root mean squared error of parallax measurements (in mcm) as a function of the decimation factor as measured on 31 randomly selected fragments of a training stereo air photograph analyzed by a professional human operator.

One can see from these data that 4x- and even 5x-decimation/interpolation of one of two images of a stereo pair do not dramatically increase the measurement error. With the increase of the decimation order, RMS error grows according to a parabolic law. These experiments showed also that, after 7x-decimation/interpolations, localization failures appear, and the probability of failures grows with further increase of the decimation factor very rapidly from probability of failures 5% at 7x-decimation to 21% at 8x-decimation. All this is in a good correspondence with the theory of localization accuracy of image correlators ([8]), although the data were obtained for a human operator.

An extended series of experiments on human stereo vision tolerance to image blur is described in what follows. In these experiments, measured were stereopsis threshold and parallax measurement accuracy as functions of the degree of blur of one of the two stereo images.

In the experiments on the stereopsis threshold, synthetic stereo images with different parallax were displayed for viewers who were requested to detect a 3D target by indicating its position using computer generated cross cursor. Computer program registered the correct detection if the cursor was placed within the target object. For each degree of blur, from no blur to maximal blur, the target in form of a circle was randomly placed within the image and its parallax was changed from low to higher values until viewer detects the 3D target after which a new session with another blur started. Both random dot and texture test images were used in the experiments.

In the experiments on the accuracy of parallax measurement, a test target with a moderate parallax was randomly placed within the image and the viewers were requested to place a computer generated cross mark to the center of the target as soon as it was detected. Coordinates of the mark were compared by the computer program with actual coordinates of the target and the localization error was calculated. Each measurement was repeated, for the same image blur, several tens of times for statistical estimation of the localization error mean value and standard deviation that were used to characterize the parallax measurement accuracy. Note that parallax values in these experiments were taken to be higher than the threshold values found in the previous experiments. This was done because in localization accuracy experiments it is necessary to see the 3D stimulus clearly.

As it was already mentioned, two methods of image blur were used: approximated ideal low pass filter, and blurring using low pass filters used in wavelet image decomposition.

Approximated low pass filters were implemented in the domain of Discrete Fourier Transform. In order to avoid image oscillations caused by sharp cut-off of the ideal low pass filter, the rectangular frequency response of the ideal low pass filter was smoothed to make the filter cut-off less sharp. The filters were specified by the fraction, from 1 to 0.1 with a step of 0.1, of image bandwidth it passes through. Wavelet low pass filters were specified by the type of the wavelet used and by the resolution level with the resolution level 1 corresponding to the original image resolution and the resolution level $\boldsymbol{n}$ corresponding to resolution $\mathbf{2^{-(n-1)}}$ – th of the original resolution. In both cases, the degree of blur specified either by filter bandwidth or wavelet decomposition resolution level can be directly translated into the reduction of the number of samples required to generate the blurred image.

In addition to the detection threshold and localization error, time delay between prompting viewers to respond and their response was also measured in all experiments.

Typical results of the experiments are summarized in Figs. 5-7. Both detection threshold and standard deviation of the localization error in these figures as well as the depth values indicated in Figs. 6 and 7 are given in terms of the parallax values introduced for test object, the values being measured in units of inter-pixel distance, such that threshold or error equal to 1 correspond to the parallax in one sampling interval.

Figs. 5 a,b, represent data obtained in experiments on stereopsis threshold for two test images, random dots and color random patches carried out with image blurring using an approximated low pass filter. The blur factor in these graphs indicates the fraction of the image base band left by the low pass filter: blur factor one corresponds to no low pass filtering, blur factor two corresponds to low pass filtering to half of the base band, three - low pass filtering to one third of the base band, and so on. For random dots test image, the results of 10 experiments for one viewer and their mean and median are shown to illustrate the typical data spread in experiments with one viewer. For random patches image, average data for 3 viewers are shown along with their mean and median to illustrate spread of averaged data for several viewers. Figs. 5, c), d) represent stereopsis threshold data obtained for random dot test image using for image blur Haar and Daubechis-1 wavelet low pass filters, correspondingly. In these figures, “Scale index” specifies the scale in wavelet multi-resolution decomposition: scale index one corresponds to the initial resolution, scale index two corresponds to half of the initial resolution, scale index three corresponds to quarter of the initial resolution, four – to one eight of the initial resolution, and so on.

Fig. 6 represents results of evaluating 3D target localization accuracy obtained, using approximated low pass filter, for different test images, different target depth and stereoscope and anaglyph methods of image display, data obtained for stereoscope and for anaglyph as the viewing device are shown in the left column and in the right column of the figure, correspondingly.

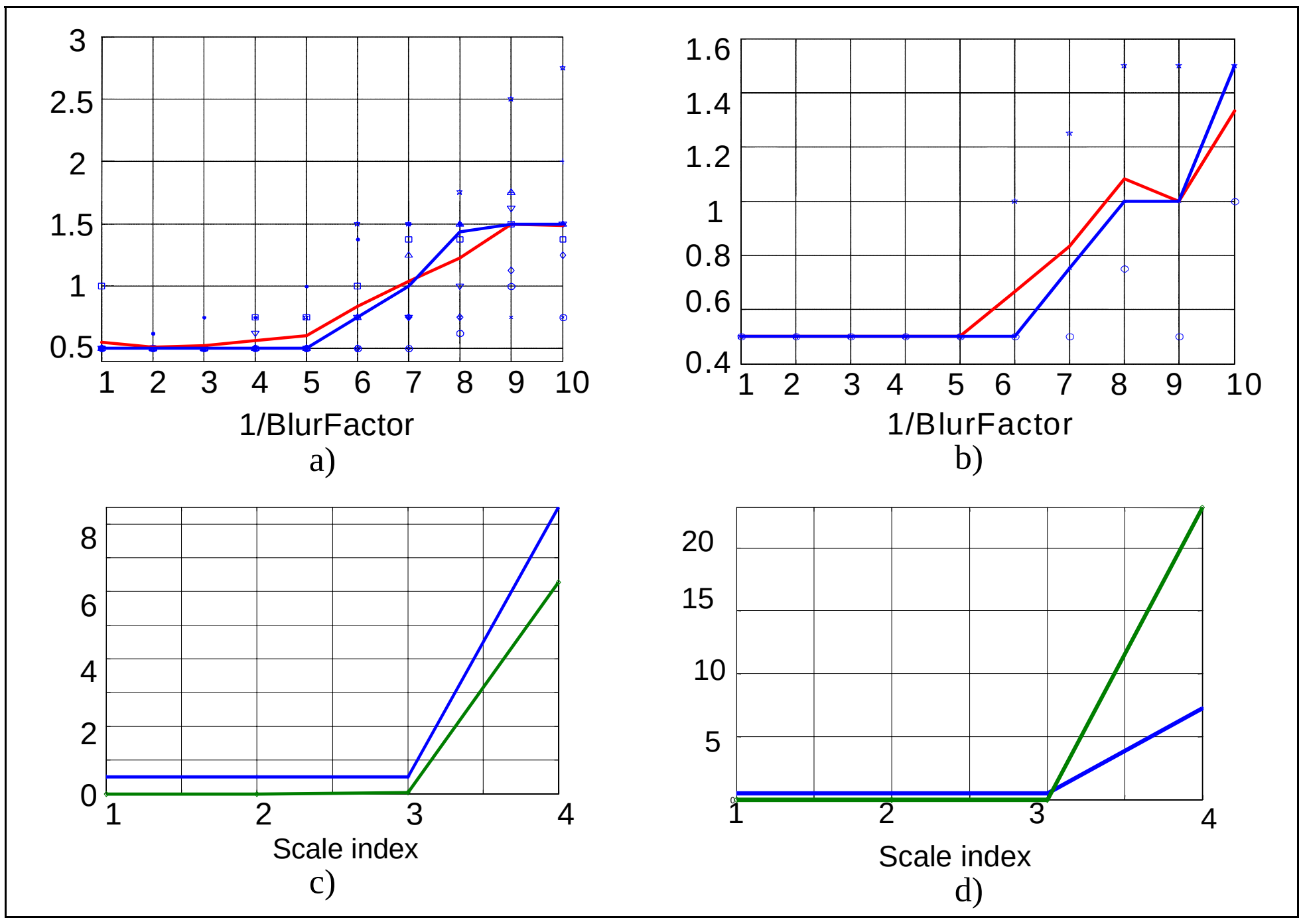

Fig.5. Stereopsis threshold as a function of image blur: a) test image “random dots”, case “ideal low pass filter , 10 experiments (dots) with one viewer and data average (red) and median (blue); b) - test image “Color random patches”, case “low pass filter, 3 viewers (dots) and data average (red) and median (blue); c) - test image “random dots”, “Haar” wavelet low pass filter”; d) - test image “random dots”, “Daubechis -1 wavelet low pass filter. In d) and c), blue curves represent mean values of stereopsis threshold measurements and green curves represents standard deviation of the measurements

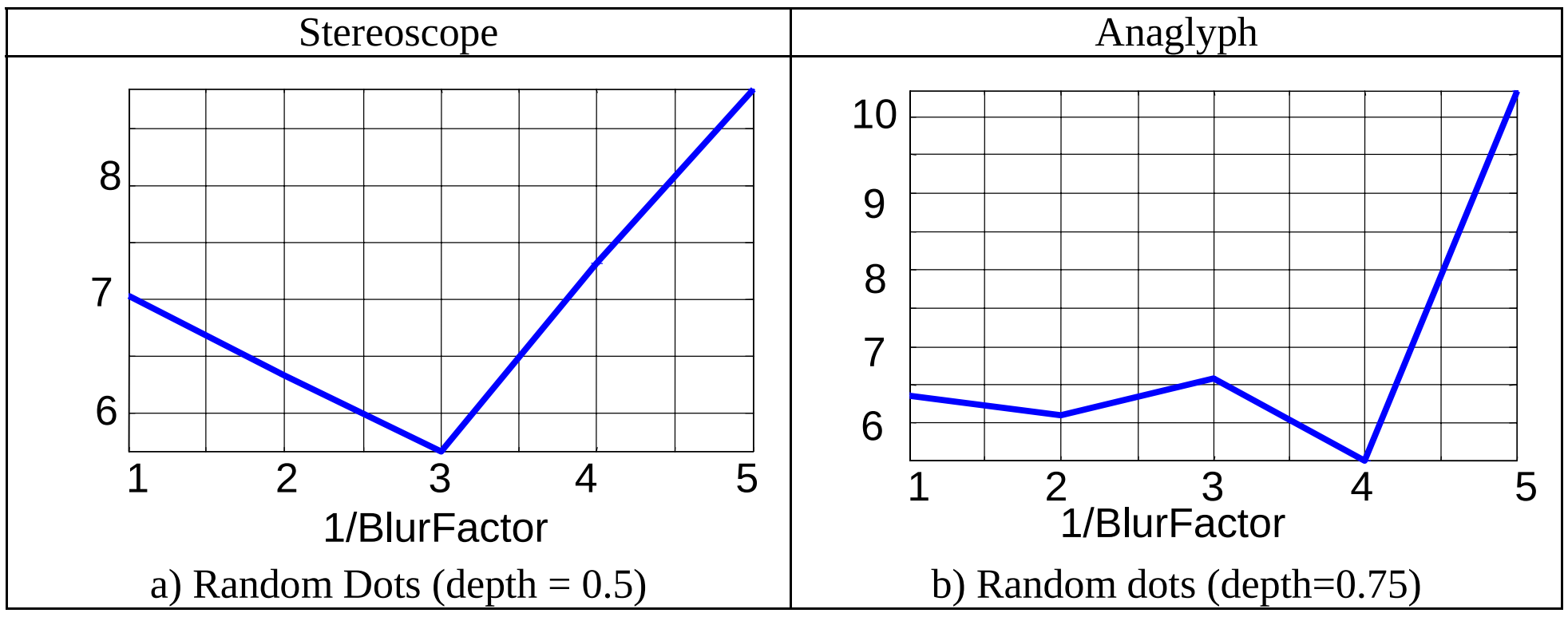

Fig. 6, a-b. Mean square of 3D target localization error as a function of image blur using approximated low pass filter

| Stereoscope | Anaglyph |
|---|---|
| 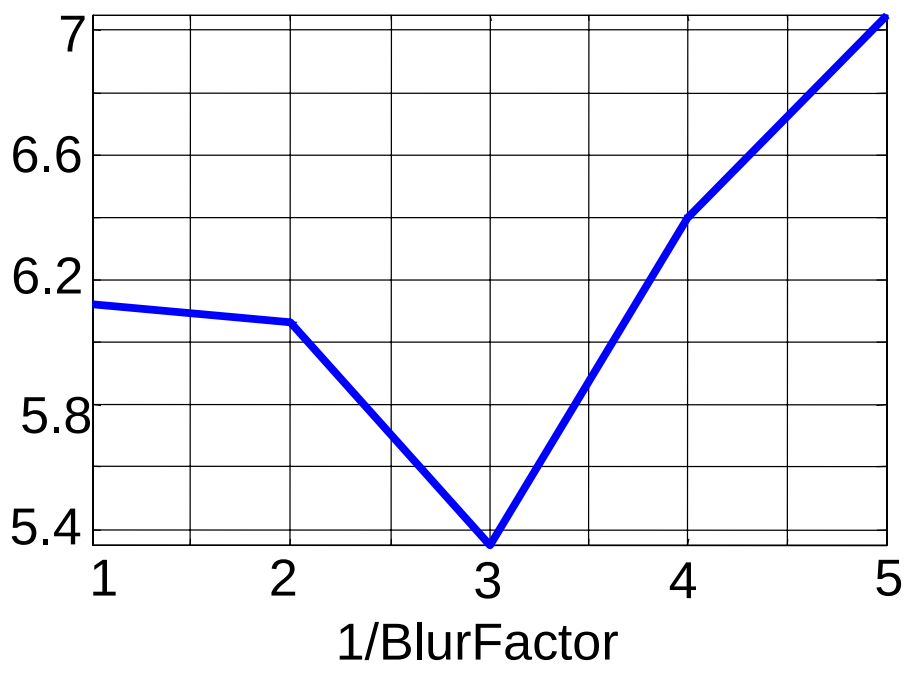 c) Piece Wise Noisy BW (depth 1.5) | 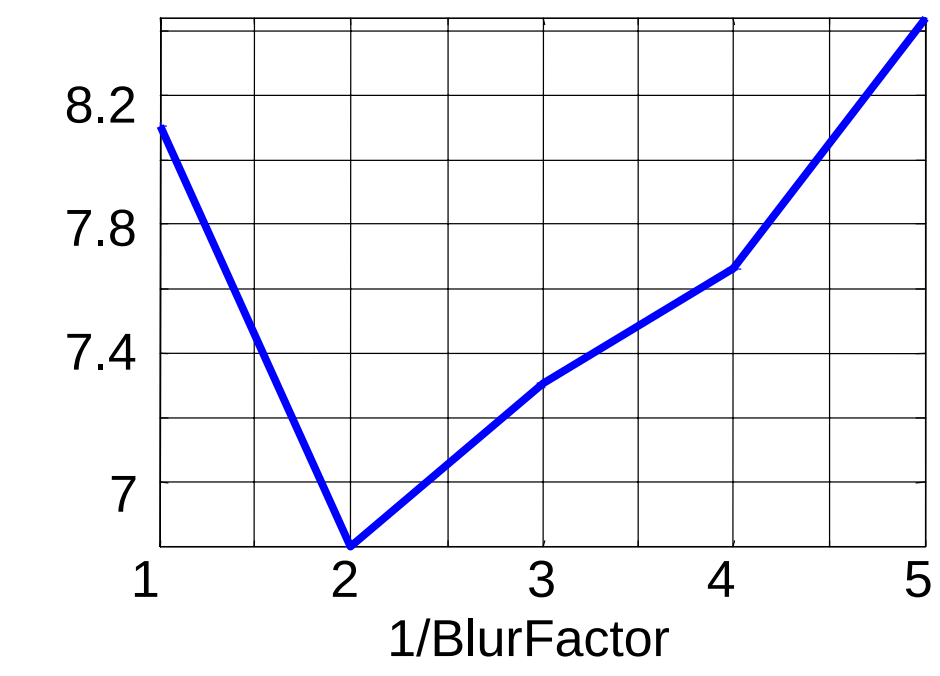 d) Piece Wise BW (depth 1.5) |
| 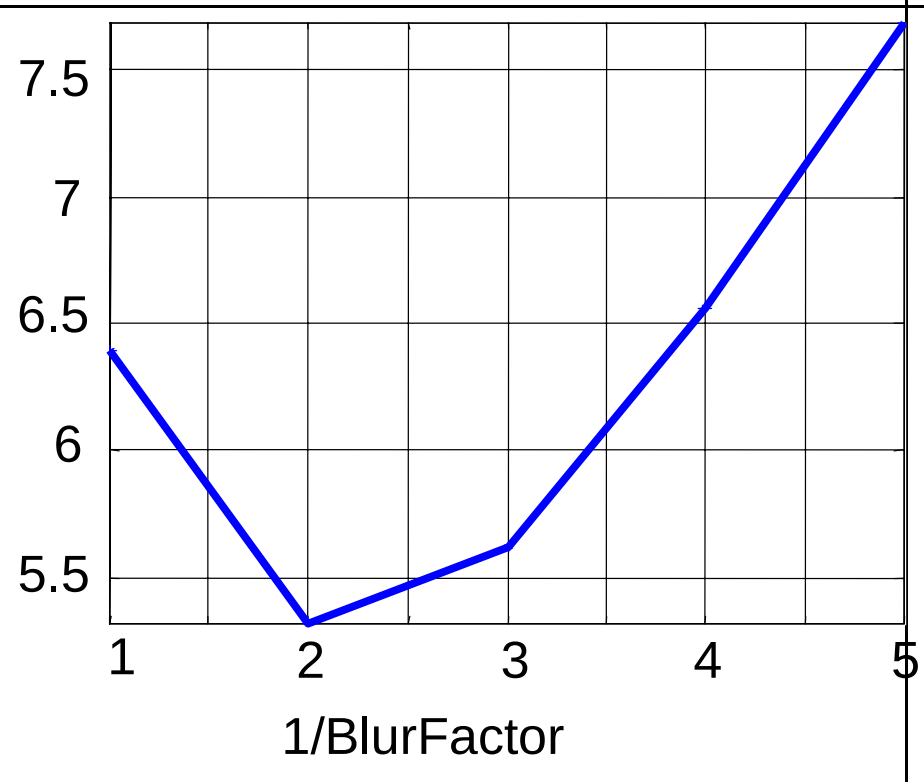 e) Textile (depth=1.75) | 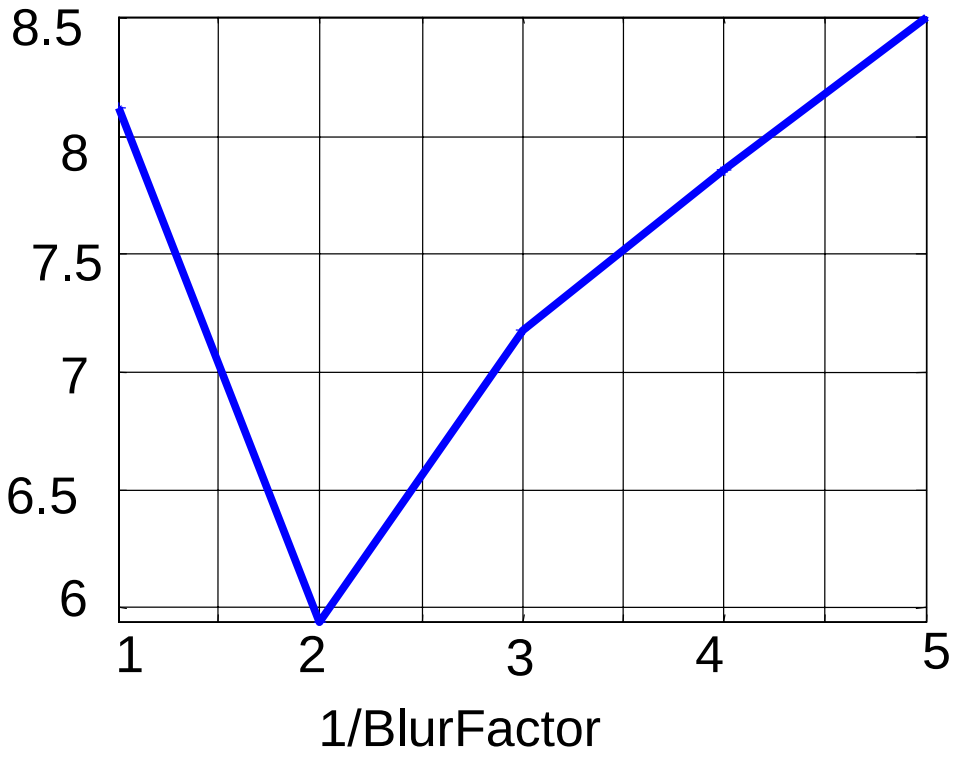 f) Textile (30 sights, depth=1) |
| 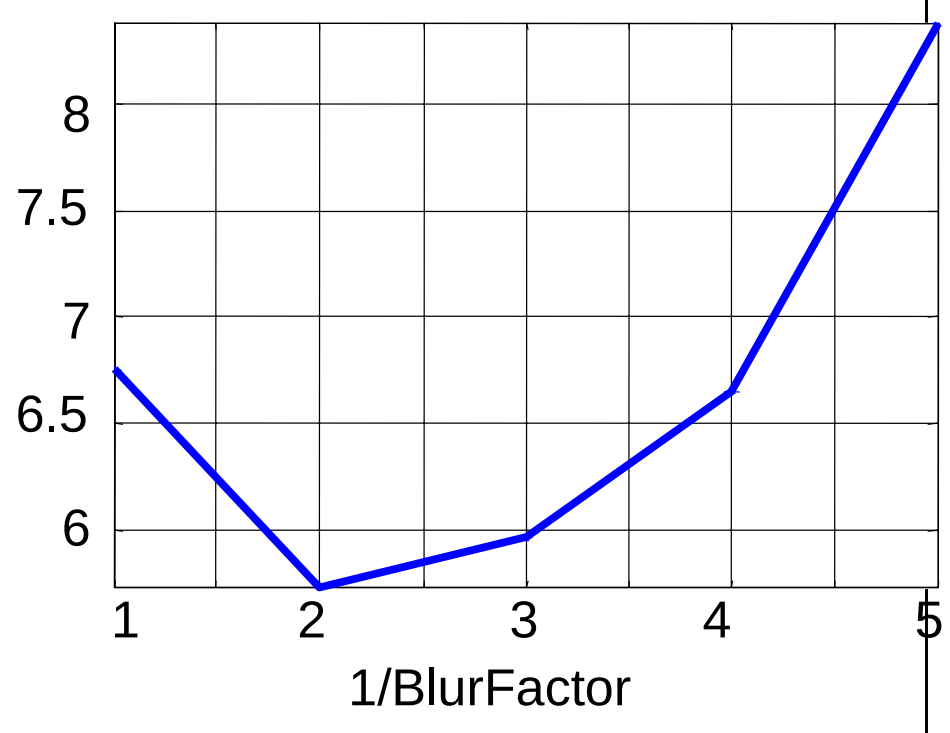 g) Texture 2 (30 sights, depth=0.5) | 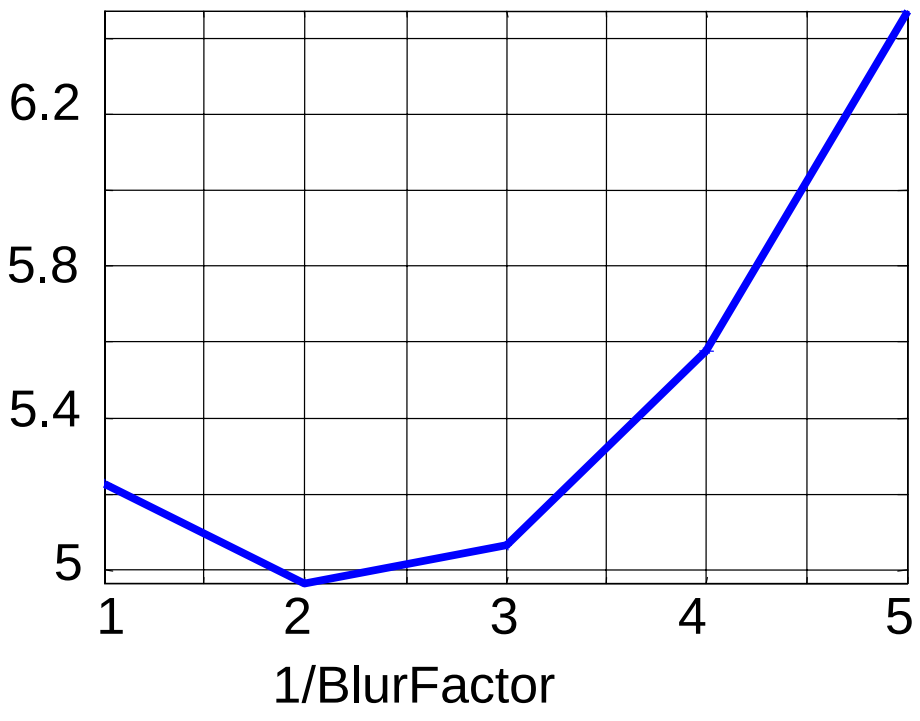 h) Texture 2 (30 sights, depth=1..5) |

Fig. 6, e-f. Mean square of 3D target localization error as a function of image blur using approximated low pass filter as a function of image blur using the approximated low pass filter.

Fig. 7 represents the threshold data obtained for random dots test image using Daubechis-1 wavelet low pass filters and with anaglyph method of image display.

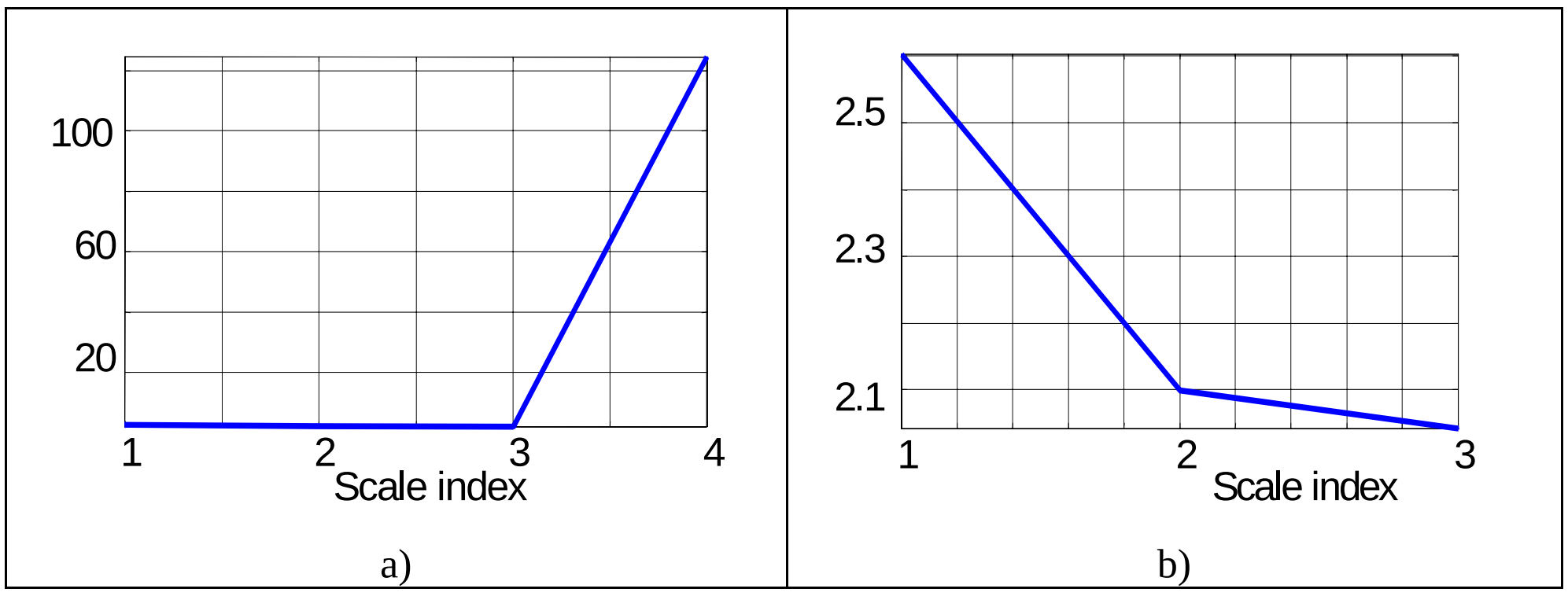

Fig 7. Mean square of 3D target localization error as a function of image blur using Daubechis-1 wavelet low pass filter, depth = 4: a) – all data; b) – data for the first three scales

## 5. Tolerance of 3D color vision to loss of color saturation

Experiments with color saturation were aimed at the determination of the threshold in loss of color saturation of one of two stereo images for which color degradations are not visible when stereo images are viewed. In experiments, $\boldsymbol{R}$, $\boldsymbol{G}$, and $\boldsymbol{B}$ color components of one of two stereoscopic images were modified in the following way:

$$R_{mod} = \overline{RGB} + g\left(R - \overline{RGB}\right),\quad G_{mod} = \overline{RGB} + g\left(G - \overline{RGB}\right),\quad B_{mod} = \overline{RGB} + g\left(B - \overline{RGB}\right).$$

where $R_{mod}$, $G_{mod}$ and $B_{mod}$ are modified color components,

$$\overline{RGB} = (R + G + B)/3$$

and $g = 0 \div 1$ is a color saturation parameter. When **g=1**, the image is unchanged; when $g = 0$, the color image is converted to a single component black and white one. Fig. 8 shows an example of color stereo images with one image of low saturation.

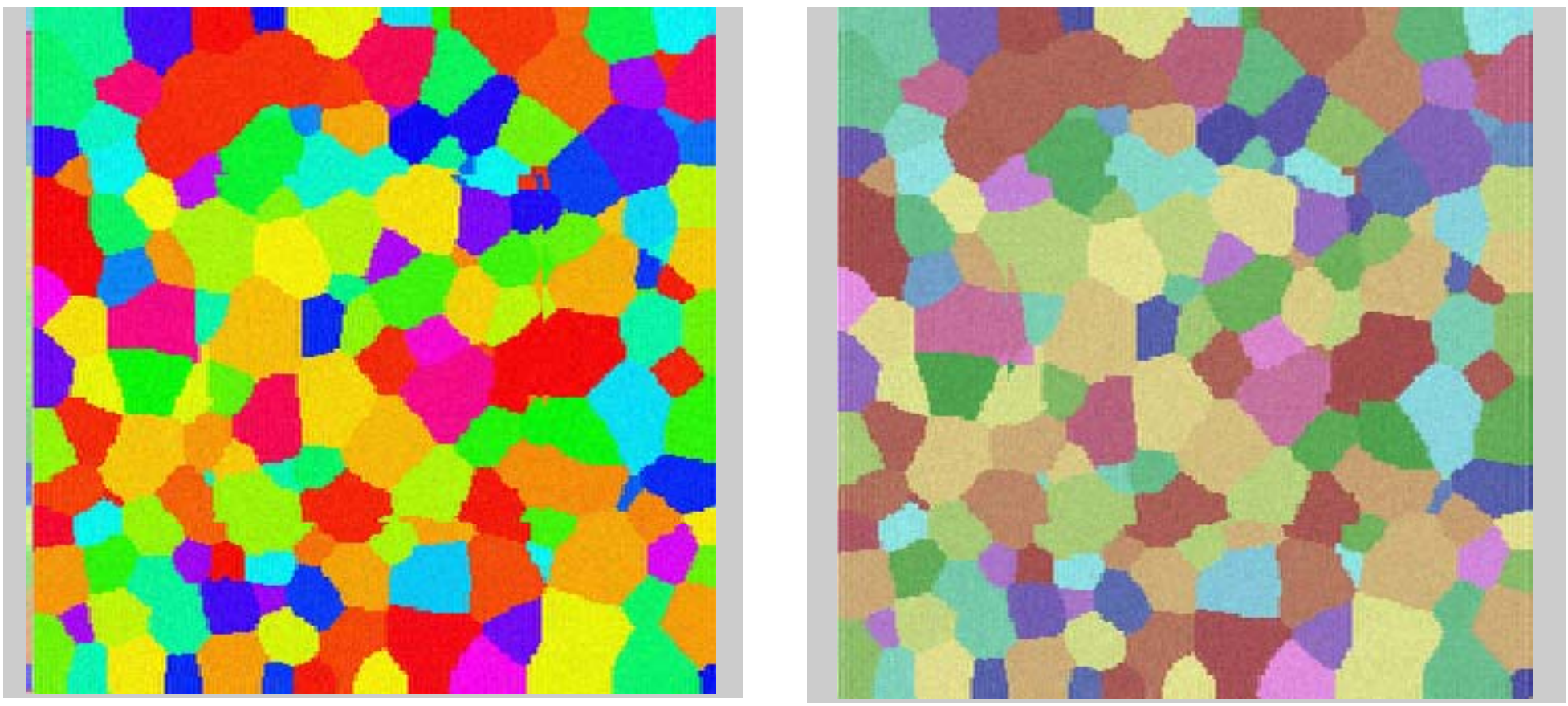

Left image Right image; saturation 0.3

Fig. 8.An example of color stereo images with one image of low saturation.

In the experiments two pairs of color stereoscopic images were simultaneously presented to the viewers for visual comparison: full color original stereoscopic images and stereoscopic images in which one of the images was subjected to color saturation with a certain saturation parameter. The experiments were aimed at determination of threshold values of visual detection of color loss of one pair of stereoscopic images as compared to the reference pair when saturation factor was gradually reduced from 1 to the level at which the viewer detects the losses and then from this level back to one till the viewer again acknowledges identity of the two pair of images. The results obtained with 5 viewers for computer generated and real life test images are presented in Fig. 9.

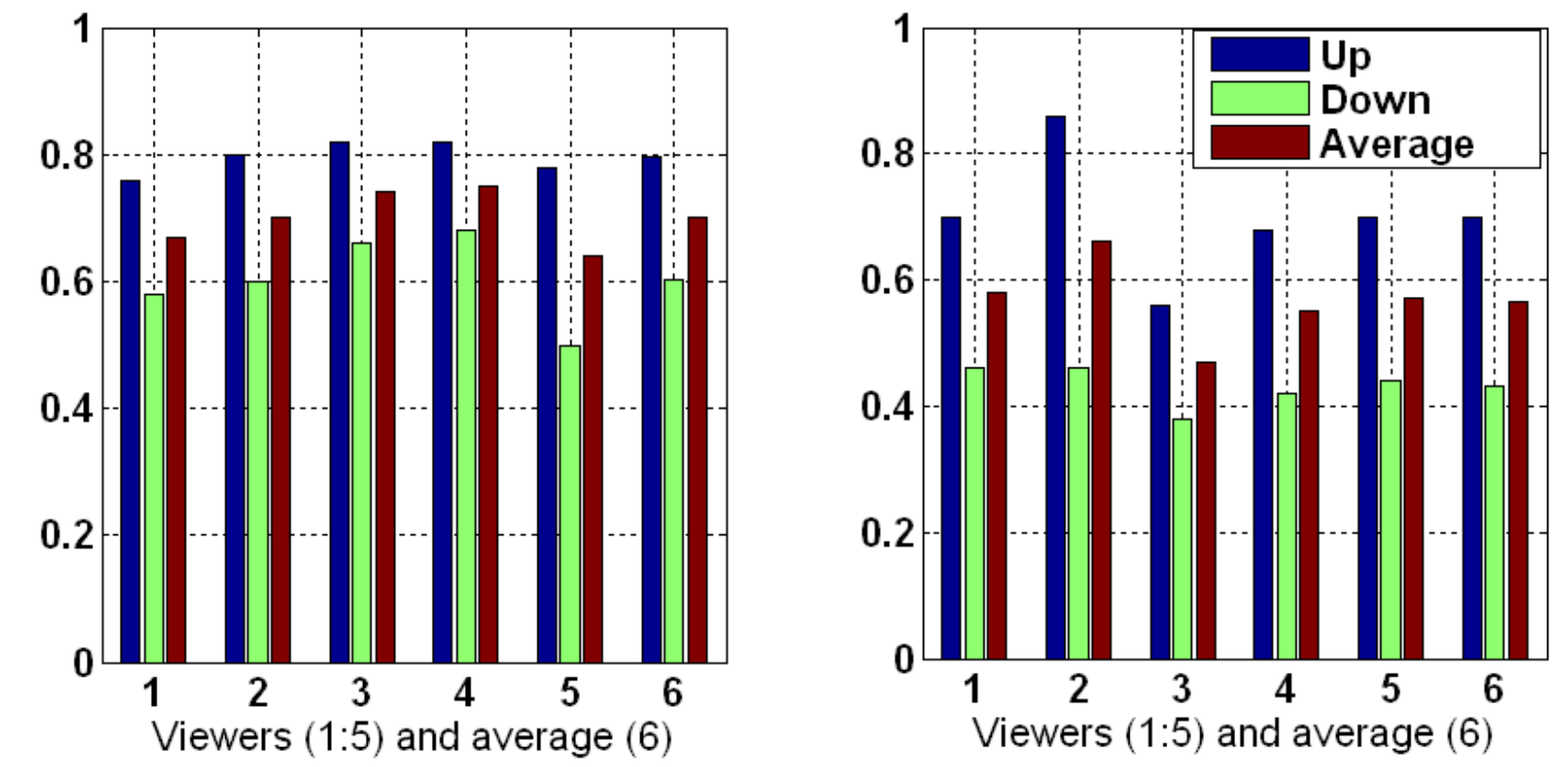

Fig. 9. Results on color saturation (left: random patches image, Fig. 1, f; right: real life image, Fig. 1, g,h)

## 6. Discussion and conclusion

The data obtained in the reported experiments with different test images and different viewers were reasonably consistent and stable. They definitely show that stereo images do have very substantial redundancy that can be easily exploited in the design of 3D visual communication system. Specifically, the following results can be formulated.

- Stereoscope and anaglyph data for both stereopsis threshold and accuracy are practically identical which means that these two methods of 3D image visualization are equivalent in terms of 3D visual quality they provide.
- Both threshold and target localization accuracy do not substantially suffer from blur of one of two images up to the blur that corresponds to 5x5 to 7x7 times reduction in the number of pixels of one of two stereo images. For blur factor larger then 1/8-1/10, rapid growth of the stereopsis threshold and of the localization accuracy was observed. These results are consistent with the theory of target location ([5,8]).
- 3D target localization accuracy grows with the target depth, which one also would expect from the point of view of the target localization theory.
- For both, stereopsis threshold and 3D target localization accuracy an interesting phenomenon of improvement visual performance with blur of one of two images was very frequently observed. In the cases where improvement in threshold level or in the localization accuracy was not observed, it was observed as a decrease in the time delay between the prompt for viewer to respond and the viewer’s response. This phenomenon does not seem paradoxical or contradictory to the theory of target location. It may have rational explanations. One of possible explanations is that, in stereopsis, one of two eyes is “leading” and provides the brain with a sharp image while the other serves solely for measuring the parallax for 3D perception. Moderate blur of one of the two images eases for the brain selection of the “leading” eye.
- Losses of color in one of two color stereoscopic images are not visually perceived for saturation as low as 0.5 – 0.7.

In conclusion, note that the reported experiments were not directly aimed at studying of mechanisms of human 3D and color visual perception. However, the authors believe that they cast additional light on the problem and motivate further research and dedicated experiments.

**Acknowledgement**

We would like to thank Prof. M. J. Yzuel for useful discussions. This work has been partially financed by Ministerio de Ciencia y Tecnología, under project BFM2003-006273-C02-01, and by the European Community project CTB556-01-4175. Manuel Espinola acknowledges the grant received by this European project. L. Yaroslavsky acknowledges the grant SAB2003-0053 by the Spainish Ministry of Education, Culture and Sport. I. Ideses acknowledges the PhD fellowship of Tel-Aviv University.